# Nanomolding single-crystalline CoIn$_3$ and RhIn$_3$ nanowires

Nghiep Khoan Duong,[1] Christian D. Multunas,[2,3] Thomas Whoriskey,[4] Mehrdad T. Kiani,[5] Shanta R. Saha,[6] Quynh P. Sam,[5] Han Wang,[5] Satya Kushwaha,[4,7] Johnpierre Paglione,[6] Ravishankar Sundararaman,[2] and Judy J. Cha[5]

[1] Department of Physics, Cornell University, Ithaca, NY 14853, USA
[2] Department of Materials Science and Engineering, Rensselaer Polytechnic Institute, Troy, NY 12180, USA
[3] Department of Physics, Fairfield University, Fairfield, CT 06430, USA
[4] Institute for Quantum Matter, William H. Miller III Department of Physics and Astronomy, The Johns Hopkins University, Baltimore, MD 21218, USA
[5] Department of Materials Science and Engineering, Cornell University, Ithaca, NY 14853, USA
[6] Department of Physics, Quantum Materials Center, University of Maryland, College Park, MD 20742, USA
[7] Department of Chemistry, The Johns Hopkins University, Baltimore, MD 21218, USA

Intermetallic compounds containing transition metals and group III-V metals tend to possess strong correlations and high catalytic activities, both of which can be enhanced via reduced dimensionality. Nanostructuring is an effective approach to explore this possibility, yet the synthesis of nanostructured intermetallics is challenging due to vast differences in melting points and vapor pressures of the constituent elements. In this work, we demonstrate that this challenge can be overcome with thermomechanical nanomolding (TMNM), exemplified by the synthesis of intermetallic CoIn$_3$ and RhIn$_3$ nanowires. We show that TMNM successfully extrudes single-crystalline nanowires of these compounds down to the 20 nm diameter range, and the nanowires remain metallic with resistivity values higher than calculated bulk resistivity. We discuss possible effects of surface roughness scattering, vacancy-induced scattering, and surface oxidation, on the measured resistivities of the nanowires. For CoIn$_3$ nanowires, the measured resistivity values are the first reported values for this compound.

A perennial challenge in nanomaterials synthesis is the challenge of combining elements with disparate vapor pressures, diffusivities, and melting points to form high-quality, single-crystalline nanostructures. Nanomaterials, such as nanobelts or nanowires, are often synthesized via open-environment growth techniques, such as chemical vapor deposition or vapor-liquid-solid growth, where source materials are introduced as vapors, transported via carrier gas in horizontal tubes, and deposited onto a substrate. This does not always enable elements to come together to form a single crystalline structure the way bulk synthesis methods allow, especially when there is a large difference in the vapor pressures and melting points of the constituent elements. Too high a vapor pressure difference, for example, can cause self-decomposition, as is the case for As in InAs$_{1-x}$Sb$_x$ nanowire growths, thus limiting their composition range [1]. On the other hand, the low vapor pressure of Sb, relative to those of In and Ga, can result in the formation of unwanted Sb crystallites, as well as diffusion-altering surfactant and memory effects [2], [3], leading to uncontrolled stoichiometry [4] and complicated reaction pathways [5] for the growths of indium and gallium antimonide nanowires and epitaxial layers.

Fabricating intermetallic nanowires comprising elements with disparate melting points, however, is potentially rewarding. Intermetallic compounds comprising transition-metals (such as Fe, Ni, Co, Rh, Ir, Ru, etc.) and group III-V metals (such as In, Ga, or Sn) give rise to strongly correlated compounds with interesting properties. These include flatband kagome metals (Ni$_3$In, CoSn, FeSn, etc.) [6], [7] and intermetallic superconductors (FeGa$_3$, RuIn$_3$) [8], [9]. Such compounds also make good catalysts, notably for selective hydrogenation (CoIn$_3$, CoGa$_3$) [10], water-splitting (IrIn$_2$) [11], or lithium deposition on lithium-based batteries (CoIn$_3$) [12]. Nanostructuring provides an opportunity to study strong electron correlations under reduced dimensionality [13] and to improve catalytic properties via the maximization of active sites [14].

One approach to overcome the synthesis challenge of intermetallic nanowires is to employ closed-environment synthesis techniques, where atoms are forced together to form nanostructures in the solid phase. This concept is realized in thermomechanical nanomolding (TMNM), in which interfacial diffusion pushes atoms from a bulk feedstock into a mold with nanoscale pores, enabling the extrusion of nanowires therein [15], [16], [17], [18], [19], [20]. This approach has been successful in synthesizing single-crystalline intermetallic nanowires with high aspect ratios and well-controlled diameters [15], [16], and can be an ideal approach to address the synthesis challenge of fabricating nanowires of constituent elements with disparate melting points.

In this work, we report the synthesis of single-crystalline $CoIn_3$ and $RhIn_3$ nanowires using TMNM. $CoIn_3$ and $RhIn_3$ crystallize into the tetragonal, $FeGa_3$-type crystal structure, with space group $P4_2/mnm$ (Fig. 1 (a) and (b)). The presence of Indium with its low melting point (156.6 °C) and high vapor pressures, alongside Cobalt or Rhodium, which have high melting points (1495 °C and 1964 °C, respectively), makes these compounds good candidates for testing the viability of TMNM in overcoming vast differences in melting points and high vapor pressures for nanowire synthesis. Furthermore, $CoIn_3$ and $RhIn_3$ are predicted to be topological Dirac semimetals [21], with $RhIn_3$ experimentally shown to exhibit a nontrivial Berry phase [22]. $CoIn_3$ also possesses desirable catalytic properties, having been shown to promote homogeneous lithium deposition in lithium batteries [12], as well as high hydrogenation selectivity in the production of unsaturated alcohols [10]. Fabricating nanowires of these intermetallic compounds would be beneficial for the studies of their predicted topological states as well as for the enhancement of their catalytic properties.

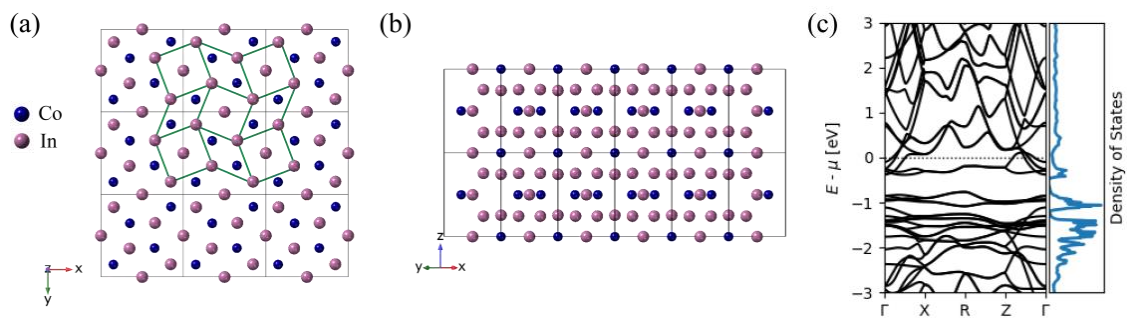

**FIG. 1**. **Crystal structure and electronic band structure of $CoIn_3$**. (a) In the (001) projection a square-net motif (green lines) is formed by Indium atoms and surrounded by rhombic prisms with pairs of Co atoms on each side of an Indium-square. (b) The (110) projection shows another motif of atomic arrangements in $CoIn_3$. (c) Calculated electronic band structure of $CoIn_3$ reveals a small density of state at the Fermi level, in proximity to a semiconducting band gap.

To that end, we have employed TMNM - or nanomolding - to synthesize $CoIn_3$ and $RhIn_3$ nanowires. The schematic in Fig. 2(a) depicts the general working principle of nanomolding: a bulk polycrystalline feedstock is first planarized and polished to mirror finish and then pressed against a nanoporous mold made of anodic aluminum oxide (AAO) at elevated pressure and temperature [15], [19]. This process enables the atoms from the feedstock to creep up along the nanopores by interfacial diffusion to form nanowires. The length of the wires can be tuned by varying the molding temperature, the pressure applied, or the duration of the molding process, and their diameters can be tuned by selecting different pore sizes for the AAO mold. As-molded wires can be isolated from the AAO mold by wet chemical etching (details in Supplementary Information) followed by sonication for further characterization.

Using an indium-rich, polycrystalline Co-In alloy as the bulk feedstock and an applied pressure of approximately 150 MPa at 350 °C for one hour, we extrude the Co-In alloy into the nanowire shape, as captured by scanning transmission electron microscopy (STEM). Figure 2(c) shows a cross-section high-angle annular dark field (HAADF) STEM image of the nanomolded sample, prepared by focused ion beam (FIB), where the extruded nanowires are seen as nanopillars standing on top of the bulk feedstock. The compositional map (Fig. 2(d)), acquired by STEM electron energy loss spectroscopy (EELS), clearly shows the diffusion of Co and In from the feedstock into the pores to form nanowires. The wires have diameters averaging 40 nm, consistent with the pore size of the mold used in the experiment. Scanning electron microscope (SEM) imaging further shows that nanomolded wires can have lengths exceeding 10 $\mu m$ (inset, Fig. 2(b)). Energy dispersive X-ray (EDX) spectra, acquired from more than 20 nanowires by transmission electron microscopy (TEM)-EDX, indicate an average stoichiometric Co:In ratio of 1:2.8, which is close to $CoIn_3$.

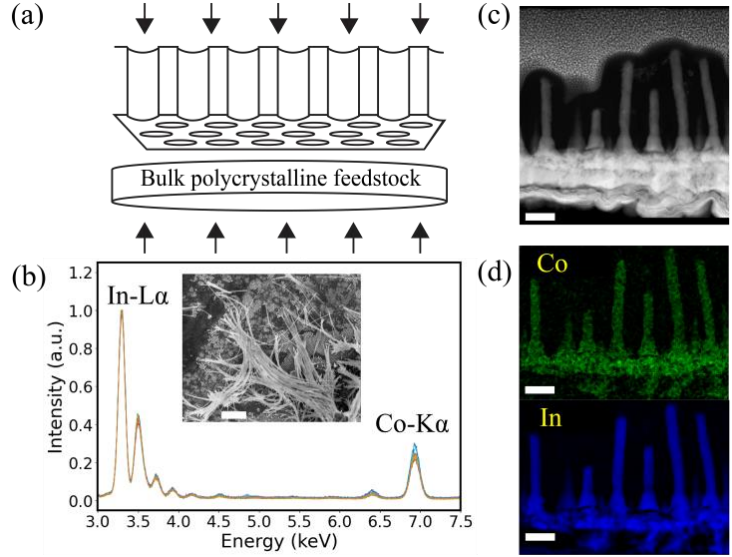

**FIG. 2. Nanomolding of $CoIn_3$ nanowires**. (a) Nanomolding schematic: a bulk feedstock is pressed against a nanoporous mold at elevated temperature under a pressure typically in the range of 100-150 MPa. After 1 hour, the bulk feedstock creeps into the mold by interfacial diffusion, forming nanowires inside the pores. The wires can then be isolated by sonication. (b) TEM-EDX spectra of molded nanowires indicate a Co:In ratio of 1:2.8, while SEM imaging (inset) shows that the wires could be up to 10 $\mu$m long (scale bar: 2 $\mu$m). (c) FIB-liftout of a portion of molded wires, imaged by HAADF-STEM, and (d) STEM-EELS chemical compositional maps of the region in (c), showing Cobalt and Indium filling the pores of the mold to form $CoIn_3$ nanowires (scale bars: 100 nm).

Atomic-resolution HAADF-STEM imaging of two representative nanowires shows the two crystallographic directions of $CoIn_3$ in their plan views (Fig. 3). Figure 3(b) shows the (001)-projection of $CoIn_3$, matching the schematic shown in Fig. 1(a), and the [130] growth direction is determined by the fast Fourier transformed (FFT) image as well as the simulated diffraction pattern (Fig. 3(c), (d)). Fig. 3(f)-(h) show another growth direction of [221] as determined from the (110)-projection HAADF STEM image of $CoIn_3$ (Fig. 3(f)), matching the schematic shown in Fig. 1(b). Single crystalline nanowires are also molded from AAO molds with 80 nm and 20 nm pore sizes, producing wires with approximately 80 nm and 20 nm diameters, respectively (Supplementary Fig. S1). Supplementary Fig. S2 shows the characterization for molded $RhIn_3$ nanowires.

We have thus demonstrated that nanomolding successfully fabricates $CoIn_3$ and $RhIn_3$ nanowires in the single crystalline form with variable diameters. These results enable measurements of the electronic properties in nanowires in comparison to bulk properties, using their dimensionality as a tuning knob. To that end, we first performed Density Functional Theory (DFT) calculations to compute the bulk electronic band structures of $CoIn_3$ and $RhIn_3$, using the Perdew-Burke-Ernzerhof (PBE) exchange-correlation functionals [23] on a Γ-centered mesh of 12x12x12 k-points (details in the Supplementary Information). The electronic band structure and the associated density of states of $CoIn_3$ show a small carrier density at the Fermi level, and a gap opening at approximately 0.5 eV below the Fermi level (Fig. 1(c)). These features are also observed in the electronic structure of $RhIn_3$ (Supplementary Fig. S3). The observation of a gap opening near the Fermi level is consistent with previous theoretical studies, which identify this to be an energy gap formed by the strong *d-p* orbital interaction between the transition metal orbitals and the main-group metal orbitals [24], [25], [26] . We then computed the room-temperature bulk resistivities of $CoIn_3$ and $RhIn_3$, via a first principles evaluation of the electron-phonon scattering elements, and from there, evaluated the electron mean free paths of these compounds. These parameters are summarized in Table 1.

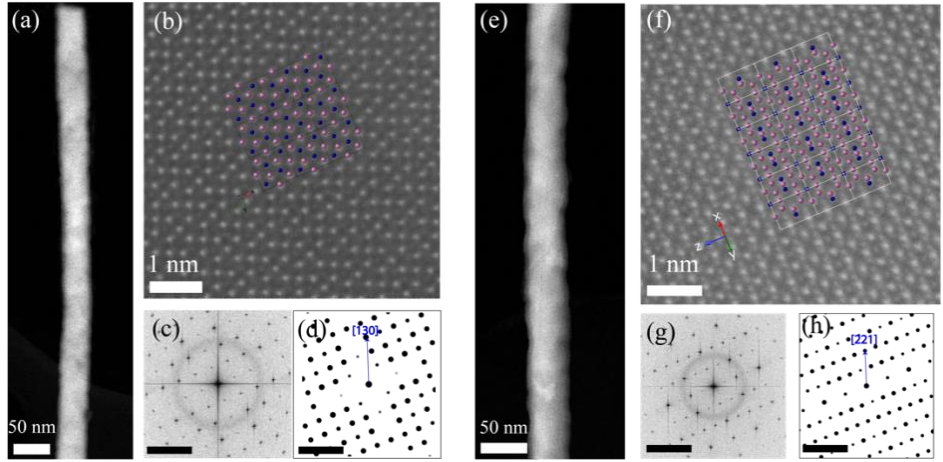

**FIG. 3. Single-crystalline CoIn$_3$ nanowires molded at 350 °C.** (a), (e) Low-magnification HAADF-STEM images of two CoIn$_3$ nanowires. (b) Atomic-resolution HAADF-STEM image of the nanowire in (a), showing the (001) projection. (c) Fourier-transformed image of (b). (d) Simulated diffraction pattern of the (001) projection, in agreement with (c) and indicating [130] growth direction. Scale bars: 5 nm$^{-1}$. (f) Atomic-resolution HAADF-STEM image of the nanowire in (e), showing the (110) projection. (g) Fourier-transformed image of (f). (h) Simulated diffraction pattern of the (110) projection, in agreement with (g) and indicating [221] growth direction. Scale bars: 5 nm$^{-1}$.

From our calculations, we expect both CoIn$_3$ and RhIn$_3$ to be metallic, with an average bulk resistivity of about 41.6 $\mu\Omega$·cm and 64.9 $\mu\Omega$·cm, respectively, at room temperature. We fabricated nanowire devices in the 4-probe configuration using standard e-beam lithography (inset, Fig. 4(a)), and measured the resistivities of individual wires (details of the device fabrication process in Supplementary Information). Fabricated with Cr/Au contacts, our devices exhibit linear $I-V$ characteristics, indicating Ohmic contacts (Fig.4(a)). Temperature-dependent resistivity measurements of CoIn$_3$ nanowires indicate that they are all metallic, down to the 20 nm diameter range (Fig. 4(b)). At room temperature, the average resistivities of CoIn$_3$ nanowires are 180.1 ± 55.5 $\mu\Omega$·cm and 529.9 ± 80.5 $\mu\Omega$·cm for nominally 40 nm diameter and 20 nm diameter wires, respectively. These values are about 4 and 13 times the calculated room-temperature bulk resistivity of CoIn$_3$. For nominal 80 nm diameter wires, the resistivity is 138.4 ± 24.4 $\mu\Omega$·cm, about 3 times the bulk resistivity. For RhIn$_3$ nanowires, a similar behavior is observed, where the average room-temperature resistivity is 347.7±95.9 $\mu\Omega$·cm for nominally 40 nm diameter wires (Supplementary Figure S4), about 5 times the calculated bulk value.

| Compound | $\rho_x$ ($\mu\Omega$.cm) | $\rho_y$ ($\mu\Omega$.cm) | $\rho_z$ ($\mu\Omega$.cm) | $\rho_{Avg}$ ($\mu\Omega$.cm) | $\lambda$ (nm) | $\tau$ (fs) | $v_F$ ($10^6$ m/s) |
|---|---|---|---|---|---|---|---|
| CoIn$_3$ | 42.66 | 42.66 | 39.49 | 41.60 | 10.1 | 25.75 | 0.39 |
| RhIn$_3$ | 72.81 | 72.81 | 49.06 | 64.89 | 2.98 | 13.54 | 0.22 |

**TABLE I.** Calculated bulk resistivities ($\rho_x$, $\rho_y$, $\rho_z$, and $\rho_{Avg}$) and other bulk parameters, namely the mean free path ($\lambda$), scattering rate ($\tau$), and Fermi velocity ($v_f$) of CoIn$_3$ and RhIn$_3$ at room temperature.

Several factors may explain the discrepancy between the measured resistivities of the nanowires and the calculated bulk resistivities. Electron transport in metallic nanowires can suffer from surface scattering, leading to resistivities higher than bulk values, even when the diameters of the wires are larger than the bulk electronic mean free path [27]. In the cases of Pt, Au, and Cu nanowires, the measured resistivities could be up to 26 times the corresponding value of the bulk [27], [28], [29]. In addition, point defects like vacancies can shift the chemical potential of the system, while at the same time introducing additional scattering centers, causing further increase to the measured resistivity. The average Co:In stoichiometric ratio of the molded nanowires is 1:2.8 (Fig. 2(b), Supplementary Figures S5-8), suggesting some degrees of In or Co vacancy, likely with Indium vacancy being more prevalent.

To quantify the effects of vacancies on the resistivity of CoIn$_3$, we performed further DFT calculations on constructed 2x2x2 supercells. Specifically, we removed an arbitrary In or Co atom to simulate a vacancy defect and recompute the scattering matrix elements in the defect supercells to calculate the resistivity in the presence of vacancy (details in the Supplementary Information). We found that Co or In-vacancy would lower the chemical potential, shifting the system to a state with increased resistivity (Fig. 4(c)). In the case of In vacancy, the chemical potential would decrease by 0.768$x$ eV, where $x$ denotes the In vacancy concentration in CoIn$_{3-x}$. Hence, for $x$ = 0.25, the chemical potential would shift downward by 0.19 eV, causing the resistivity of the material to increase to approximately 120 $\mu\Omega\cdot$cm (Fig. 4(c)). At the same time, these vacancies can also act as scattering centers, which contribute to further increase the resistivity. Taking this into consideration, our calculations find that electron scattering from vacancies would increase the resistivity further: for 1% In vacancy in the supercell, an additional 20% increase in resistivity would entail, bringing the effective resistivity to at least 144 $\mu\Omega\cdot cm$ at room temperature, in agreement with the resistivities measured in the 40 nm and 80 nm CoIn$_3$ nanowires.

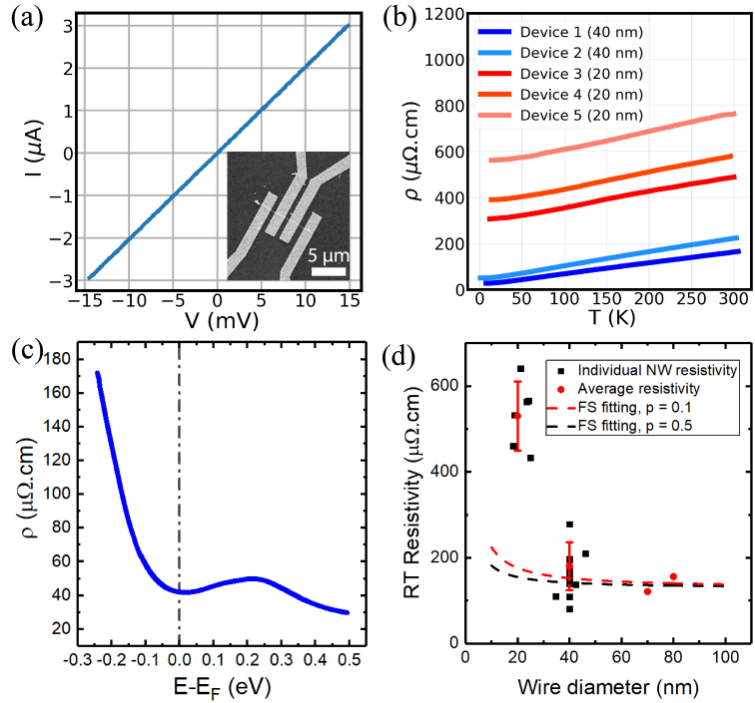

**FIG. 4. Transport properties of CoIn$_3$ nanowires.** (a) Linear $I$–$V$ characteristics exhibited by nanowire devices with 4-probe contact configuration (inset: SEM micrograph of a representative device). (b) Temperature dependent resistivity measurements show that the wires remain metallic down to 20 nm diameter. (c) Calculated room-temperature bulk resistivity of CoIn$_{3-x}$ with indium vacancy concentration, $x$. (d) Diameter-dependent, room-temperature resistivity values of molded CoIn$_3$ nanowires, in comparison to the Fuchs-Sondheimer (FS) model.

We next consider the influence of the wire diameter on the resistivity of the system. Since our nanowires are single crystalline, the effects of grain boundary scattering are ignored. As such, the dimensionality scaling of the resistivity of the system can be described by the modified Fuchs-Sondheimer formula [30], [31]:

$$\rho = \rho_0 + \frac{\rho_0 \lambda}{d} \frac{3(1-p)}{4} \tag{1}$$

where the second term represents surface boundary scattering, with $\rho_0$ being the bulk resistivity, $\lambda$ the electronic mean free path, $d$ the diameter of the nanowire, and $p$ the specularity, or the probability for specular scattering ($p = 1$ denotes an atomically smooth surface leading to perfectly specular scattering). Fig. 4(d) shows the measured room-temperature resistivities of $CoIn_3$ nanowires as a function of diameter, with the Fuchs-Sondheimer (FS) fitting for $p = 0.1$ and $p = 0.5$. While the resistivity values for nanowires with nominal 40 nm and 80 nm diameters are fitted by the FS model, the experimental resistivities for nanowires with nominal 20 nm diameter far exceed the modeled values, even for the extreme case of $p = 0.1$ (minimal specular scattering). This suggests that additional factors must be involved in influencing the nanowire resistivities at downsized dimension.

One possible factor for the discrepancy observed for 20 nm diameter wires is the degree of surface oxidation in the wires. During the device fabrication, we perform Ar plasma etching to remove a thin layer of native oxides at the nanowire surfaces; however, subsequent exposure to air before and after the deposition of metal contacts inevitably causes surface re-oxidation to the wires, which likely increases the measured resistance. Another possible contribution is contact resistance. Even though Ohmic contact was achieved, the contact resistances remain high - roughly 700-900 Ω for 40 nm diameter nanowire devices, and around 1600-2000 Ω for 20 nm diameter devices (Supplementary Figures S9, S10). As the physical contact area between the nanowire and the metal electrode decreases significantly with decreasing nanowire diameter, a sizable contribution of the experimentally measured resistance may come from the contact resistance, which increases for smaller diameter wires.

In summary, our study demonstrates the successful synthesis of $CoIn_3$ and $RhIn_3$ nanowires. Despite large differences in melting points and vapor pressures of the constituent elements, single-crystalline nanowires were formed, using the TMNM technique. From high-resolution STEM imaging, we capture the crystal structure of the wires, thus showing that they can grow in multiple directions, such as [130] and [110], among others. TEM-EDX suggests that our wires contain In vacancies, which will cause the chemical potential of the system to shift downwards by approximately 0.15 eV and thus, increase the resistivity according to DFT calculations. Other factors such as surface oxidation, surface roughness scattering, vacancy-induced scattering, and potential contact resistance will contribute to the measured resistivities, which are higher than the calculated bulk values. Nevertheless, the wires retain metallic properties down to the 20 nm diameter range. The measured resistivities of the $CoIn_3$ nanowires, to our best knowledge, are the first experimental values reported for this compound, though they represent upper bounds to the actual resistivity.

We acknowledge support from the SRC JUMP2.0 SUPREME program for device fabrication, and the Gordon and Betty Moore Foundation's EPiQS Synthesis Investigator grant (GBMF9062.01) for thermomechanical nanomolding. We also acknowledge support from the NSF DMR 2328907 for the temperature-dependent transport measurements. This work made use of the Cornell Center for Materials Research shared instrumentation facility and was performed in part at the Cornell NanoScale Facility, a member of the NNCI supported by NSF Grant No. NNCI-2025233.

[1] L. Namazi, S. G. Ghalamestani, S. Lehmann, R. R. Zamani, and K. A. Dick, "Direct nucleation, morphology and compositional tuning of InAs1- xSbx nanowires on InAs (111) B substrates," *Nanotechnology*, vol. 28, no. 16, p. 165601, 2017.

[2] A. C. Farrell, W.-J. Lee, P. Senanayake, M. A. Haddad, S. V Prikhodko, and D. L. Huffaker, "High-quality InAsSb nanowires grown by catalyst-free selective-area metal–organic chemical vapor deposition," *Nano Lett*, vol. 15, no. 10, pp. 6614–6619, 2015.

[3] B. M. Borg and L.-E. Wernersson, "Synthesis and properties of antimonide nanowires," *Nanotechnology*, vol. 24, no. 20, p. 202001, 2013.

[4] F. Zhou *et al.*, "Effect of growth base pressure on the thermoelectric properties of indium antimonide nanowires," *J Phys D Appl Phys*, vol. 43, no. 2, p. 25406, 2009.

[5]     F. Dimroth, C. Agert, and A. W. Bett, "Growth of Sb-based materials by MOVPE," *J Cryst Growth*, vol. 248, pp. 265–273, 2003.

[6]     W. R. Meier *et al.*, "Flat bands in the CoSn-type compounds," *Phys Rev B*, vol. 102, no. 7, p. 75148, 2020.

[7]     L. Ye *et al.*, "Hopping frustration-induced flat band and strange metallicity in a kagome metal," *Nat Phys*, pp. 1–5, 2024.

[8]     K. Umeo, Y. Hadano, S. Narazu, T. Onimaru, M. A. Avila, and T. Takabatake, "Ferromagnetic instability in a doped band gap semiconductor FeGa 3," *Physical Review B—Condensed Matter and Materials Physics*, vol. 86, no. 14, p. 144421, 2012.

[9]     Y. Takagiwa, K. Kitahara, Y. Matsubayashi, and K. Kimura, "Thermoelectric properties of FeGa3-type narrow-bandgap intermetallic compounds Ru (Ga, In) 3: Experimental and calculational studies," *J Appl Phys*, vol. 111, no. 12, 2012.

[10]    Y. Yang *et al.*, "Selective hydrogenation of cinnamaldehyde over Co-based intermetallic compounds derived from layered double hydroxides," *ACS Catal*, vol. 8, no. 12, pp. 11749–11760, 2018.

[11]    C. He *et al.*, "Low-Iridium-Content IrIn2 Intermetallics with an Unconventional Face-Centered Orthorhombic Phase for Efficient Overall Water Splitting," *Adv Funct Mater*, vol. 34, no. 8, p. 2311683, 2024.

[12]    Y. Wang *et al.*, "Promoting homogeneous lithium deposition by facet-specific absorption of CoIn3 for dendrite-free lithium metal anodes," *Nano Energy*, vol. 119, p. 109093, 2024.

[13]    H. A. Nilsson *et al.*, "Giant, level-dependent g factors in InSb nanowire quantum dots," *Nano Lett*, vol. 9, no. 9, pp. 3151–3156, 2009.

[14]    H. Y. Kim, M. Jun, S. H. Joo, and K. Lee, "Intermetallic Nanoarchitectures for Efficient Electrocatalysis," *ACS Nanoscience Au*, vol. 3, no. 1, pp. 28–36, 2022.

[15]    N. Liu *et al.*, "General nanomolding of ordered phases," *Phys Rev Lett*, vol. 124, no. 3, p. 36102, 2020.

[16]    N. Liu *et al.*, "Unleashing nanofabrication through thermomechanical nanomolding," *Sci Adv*, vol. 7, no. 47, p. eabi4567, 2021.

[17]    M. T. Kiani *et al.*, "Nanomolding of metastable Mo4P3," *Matter*, vol. 6, no. 6, pp. 1894–1902, 2023.

[18]    Q. P. Sam *et al.*, "Nanomolding of Two-Dimensional Materials," *ACS Nano*, vol. 18, no. 1, pp. 1110–1117, 2023.

[19]    M. T. Kiani and J. J. Cha, "Nanomolding of topological nanowires," *APL Mater*, vol. 10, no. 8, 2022.

[20]    M. T. Kiani, Q. P. Sam, Y. S. Jung, H. J. Han, and J. J. Cha, "Wafer-Scale Fabrication of 2D Nanostructures via Thermomechanical Nanomolding," *Small*, vol. 20, no. 17, p. 2307289, 2024.

[21]    M. G. Vergniory, L. Elcoro, C. Felser, N. Regnault, B. A. Bernevig, and Z. Wang, "A complete catalogue of high-quality topological materials," *Nature*, vol. 566, no. 7745, pp. 480–485, 2019.

[22]    L. An *et al.*, "Large linear magnetoresistance and nontrivial band topology in In3Rh," *Appl Phys Lett*, vol. 122, no. 20, 2023.

[23]    J. P. Perdew, K. Burke, and M. Ernzerhof, "Generalized gradient approximation made simple," *Phys Rev Lett*, vol. 77, no. 18, p. 3865, 1996.

[24]    V. Y. Verchenko and A. A. Tsirlin, "Semiconducting and Metallic Compounds within the IrIn3 Structure Type: Stability and Chemical Bonding," *Inorg Chem*, vol. 61, no. 7, pp. 3274–3280, 2022.

[25]    Y. Imai and A. Watanabe, "Electronic structures of semiconducting FeGa3, RuGa3, OsGa3, and RuIn3 with the CoGa3-or the FeGa3-type structure," *Intermetallics (Barking)*, vol. 14, no. 7, pp. 722–728, 2006.

[26]    P. Viklund, S. Lidin, P. Berastegui, and U. Häussermann, "Variations of the FeGa3 structure type in the systems CoIn3- xZnx and CoGa3- xZnx," *J Solid State Chem*, vol. 165, no. 1, pp. 100–110, 2002.

[27]    W. Gu, H. Choi, and K. K. Kim, "Universal approach to accurate resistivity measurement for a single nanowire: Theory and application," *Appl Phys Lett*, vol. 89, no. 25, 2006.

[28]    G. De Marzi, D. Iacopino, A. J. Quinn, and G. Redmond, "Probing intrinsic transport properties of single metal nanowires: direct-write contact formation using a focused ion beam," *J Appl Phys*, vol. 96, no. 6, pp. 3458–3462, 2004.

[29]    K. Biswas, Y. Qin, M. DaSilva, R. Reifenberger, and T. Sands, "Electrical properties of individual gold nanowires arrayed in a porous anodic alumina template," *physica status solidi (a)*, vol. 204, no. 9, pp. 3152–3158, 2007.

[30]  K. Fuchs, "The conductivity of thin metallic films according to the electron theory of metals," in *Mathematical Proceedings of the Cambridge Philosophical Society*, 1938, pp. 100–108.

[31]  E. H. Sondheimer, "The mean free path of electrons in metals," *Adv Phys*, vol. 50, no. 6, pp. 499–537, 2001.